\documentclass{article}

\usepackage{arxiv}

\usepackage[utf8]{inputenc} 
\usepackage[T1]{fontenc}    
\usepackage{hyperref}       
\usepackage{url}            
\usepackage{booktabs}       
\usepackage{amsfonts}       
\usepackage{nicefrac}       
\usepackage{microtype}      
\usepackage{lipsum}
\usepackage{multirow}
\usepackage{graphicx}
\newcommand{\printfnsymbol}[1]{%
  \textsuperscript{\@fnsymbol{#1}}%
}
\title{Learning Euler's Elastica Model for Medical Image Segmentation}
\author{
    Xu Chen\thanks{Equal contribution}\\
    Department of Eye\\ and Vision Science\\
    University of Liverpool
    \And
    Xiangde Luo\footnotemark[1]\\
    School of Mechanical\\ and Electrical Engineering\\
    University of Electronic Science\\ and Technology of China 
    \And
    Yitian Zhao\\
    Cixi Institute of \\Biomedical Engineering\\
    Chinese Academy of Sciences
    \And
    Shaoting Zhang\\
    School of Mechanical\\ and Electrical Engineering\\
    University of Electronic Science\\ and Technology of China 
    \And
    Guotai Wang\thanks{Corresponding authors, guotai.wang@usetc.edu.cn and yalin.zheng@liverpool.ac.uk}\\
    School of Mechanical\\ and Electrical Engineering\\
    University of Electronic Science\\ and Technology of China 
    \And
    Yalin Zheng\footnotemark[2]\\
    Department of Eye\\ and Vision Science\\
    University of Liverpool
}

\begin{document}
\maketitle

\begin{abstract}
Image segmentation is a fundamental topic in image processing and has been studied for many decades. Deep learning-based supervised segmentation models have achieved state-of-the-art performance but most of them are limited by using pixel-wise loss functions for training without geometrical constraints. Inspired by the Euler’s elastica model and recent active contour models introduced into the field of deep learning, we propose a novel active contour with elastica (ACE) loss function incorporating elastica (curvature and length) and region information as geometrically-natural constraints for the image segmentation tasks. We introduce the mean curvature i.e. the average of all principal curvatures, as a more effective image prior to represent curvature in our ACE loss function. Furthermore, based on the definition of mean curvature, we propose a fast solution to approximate the ACE loss in three-dimensional (3D) by using Laplace operators for 3D image segmentation. We evaluate our ACE loss function on four 2D and 3D natural and biomedical image datasets. Our results show that the proposed loss function outperforms other mainstream loss functions on different segmentation networks. Our source code is available at: \url{https://github.com/HiLab-git/ACELoss}.
\end{abstract}

\section{Introduction}
\label{sec:intro}
Image segmentation is a challenging problem in image processing and has been studied for many decades. Snake/active contour model (ACM) is firstly proposed by Kass M et al. \cite{kass1988snakes} that converts image segmentation problems into energy minimization problems where the energy of snake/active contours is optimised towards the object's boundaries. After that, \textit{Active contour without edge} model a.k.a \textit{Chan-Vese} (CV) model \cite{chan1999active} has been widely developed in the past two decades \cite{wang2018active, chan2000active} which can be formulated as below,
\begin{equation}\label{CVM}
E_{CV}(\phi,c_{1},c_{2}) = \int_{\Omega } \left | \nabla H_{\epsilon }(\phi) \right | + \lambda R
\end{equation}
where the first term of Eq.~\ref{CVM} is the length of the active contour. The second term is the inside and outside regions of the contour: $R =  \int_{\Omega } (c_{1} - f)^{2}H_{\epsilon }(\phi) + (c_{2} - f)^{2}(1-H_{\epsilon }(\phi))$. $\Omega \subset \mathbb{R}^{n}$ is a closed subset of the image $f$ to be segmented. $c_{1}$, $c_{2}$ are the mean value of the inside (foreground) and outside (background) regions respectively. $\phi$ is a level set function where the zero level curve represents the segmentation boundary. $H_{\epsilon }$ is a smooth approximation of the Heaviside function. $\lambda$ is a positive hyper-parameter to control the balance between the two terms. In CV model, level set method is often involved to optimize the model through solving partial differential equations (PDEs) in an iterative manner. To tackle the local minimum problem in solving CV models, Bresson X et al. proposed a \textit{fast global minimization-based active contour model} (FGM-ACM) \cite{bresson2007fast} to obtain a global minimum of the ACM with a dual formulation of the total variation (TV) norm.

Recently, Euler’s elastica model is employed in CV model for the segmentation of elongated structures by Zhu W et al \cite{zhu2013image}. This model will be denoted as CVE model hereafter for brevity. The CVE model can be expressed as the minimization of the following functional, 
\begin{equation}\label{CVEM}
E_{CVE}(\phi,c_{1},c_{2}) = \int _{\Omega} \Big[\alpha + \beta \Big(\nabla \cdot \frac{\nabla  \phi}{\left |\nabla \phi \right |}\Big)^2 \Big]\left |\nabla H_{\epsilon }(\phi)  \right |  + \lambda R
\end{equation}
where $\alpha$ and $\beta$ are positive parameters to control the trade-off between the length and curvature of the segmentation boundary. CVE model has several intrinsic features compared to the CV model: 1) introducing curvature profile, 2) preserving connectedness by connecting broken parts of segmentation object to form a meaningful segmentation object, and 3) reducing missing boundaries interpolation for tubular/curvilinear structures. However, the common challenges for CVE model are complexity in numerical schemes, high computational cost and slow to solve because it requires iteration to solve high order PDEs.

Inspired by the general idea of CVE and FGM-ACM model and recent ACMs-based work in deep learning \cite{gur2019unsupervised,chen2019learning}, our proposed active contour with elastica (ACE) loss function incorporates curvature, length and area of active contours and integrates them as geometric constraints into a deep learning model for image segmentation, where the number of parameters need to be converged are reduced, due to the benefits of supervised learning and some parameters can be treated as fixed hyper parameters to minimize our differentiable ACE loss in an end-to-end fashion. 
\begin{figure}[!t]
	\centering
	\includegraphics[width=0.9\linewidth]{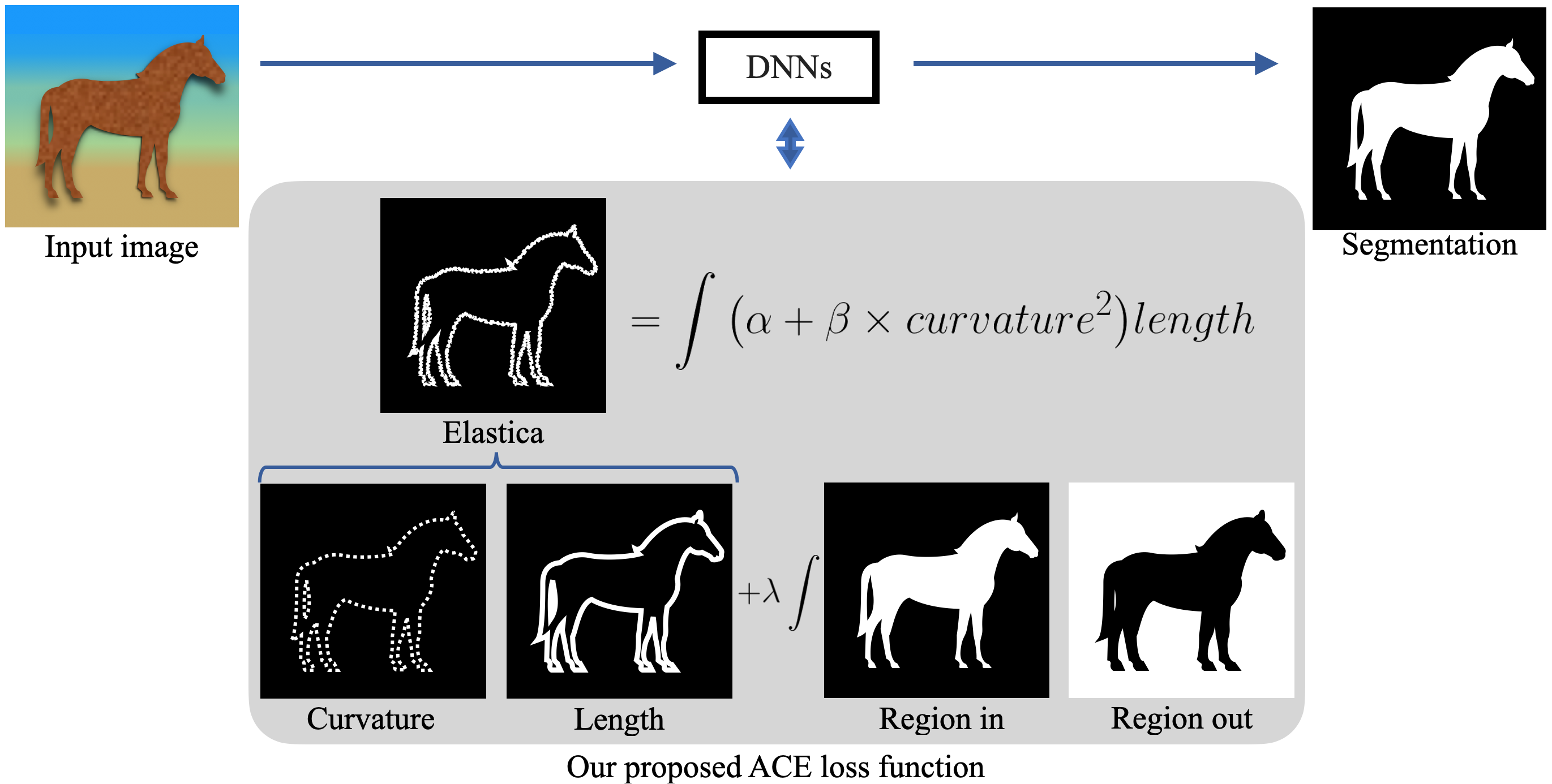}
	\caption{Overview of the proposed novel ACE loss function that integrates elastica (curvature and length) and region terms as geometrically-natural constraints for image segmentation.}
	\label{overview}
\end{figure}
\section{Methodology}\label{Method}
Our ACE loss function is defined into discrete form in the case of 3D images as follows,
\begin{equation}\label{cveloss}
\label{aceloss}
E_{ACE} (u,v) = (\alpha + \beta \bar{K}^{2} )\left | \nabla u \right | + \lambda \left |  \sum_{i=1}^{\Omega}\sum_{j=1}^{\Omega}\sum_{k=1}^{\Omega} u_{i,j,k}(c_{1} - v_{i,j,k})^{2} \right | + \lambda \left | \sum_{i=1}^{\Omega}\sum_{j=1}^{\Omega}\sum_{k=1}^{\Omega} (1-u_{i,j,k})(c_{2} - v_{i,j,k})^{2} \right |
\end{equation}
The binary ground truth mask and the predicted segmentation are denoted as $v$, $u$ : $\Omega$  $\rightarrow$ $\mathbb{R}^{3}$, respectively. $\bar{K}$ is the curvature of $u$; $c_{1}$ and $c_{2}$ are the mean intensity of the inside (foreground) and outside (background) regions respectively and can be defined as constants in advance as $ c_{1} = \mathbf{1}$ and $ c_{2} = \mathbf{0}$ \cite{chen2019learning}; $\lambda$ is usually set to 1. In our ACE loss, we define $\bar{K}$ by estimating $mean \; curvature$ \cite{taylor1992ii, gong2019weighted}, because that can be as an effective and physically-natural prior to give a more precise curvature by taking the mean of all principal curvatures \cite{gong2015spectrally}. Although $mean \; curvature$ has attractive features, minimizing $mean \; curvature$-based elastica regularization term in classical CVE model as shown in Eq. \ref{CVEM} is still far from practical given huge computational resources because it leads to a fourth-order PDE \cite{zhu2012image,gong2019weighted}, even other efficient solvers, such as fixed point method \cite{yang2014relaxed} and multi grid method \cite{brito2010multigrid}, can not reduce the computational time to a satisfactory level for real applications. The $mean \; curvature$ can be derived from Monge patch in the case of 3D images as follows,
\begin{equation}\label{h3D}
 \bar{K} = \frac{\kappa_{1}+\kappa_{2}+\kappa_{3}}{3} = \frac{\chi }{\sqrt{1+u_{x}^2 + u_{y}^2+ u_{z}^2}}
\end{equation}
where $\kappa_{1},\kappa_{2},\kappa_{3}$ are the three principal curvatures (eigenvalues) at a given point on a surface; $\chi=u_{xx}(1+u^{2}_{y}+u^{2}_{z})+u_{yy}(1+u^{2}_{x}+u^{2}_{z})+u_{zz}(1+u^{2}_{x}+u^{2}_{y})-2(u_{x}u_{y}u_{xy} + u_{x}u_{z}u_{xz} + u_{y}u_{z}u_{yz})$; $x$, $y$ and $z$ are three different direction respectively in the case of 3D images. In the case of 2D images, $mean \; curvature$ can be derived as shown in Eq.~\ref{cve2D},
\begin{equation}\label{cve2D}
 \bar{K}_{2D} = \frac{(1+u^{2}_{x})u_{yy}+(1+u^{2}_{y})u_{xx}-2u_{x}u_{y}u_{xy}}{2(1+u^{2}_{x}+u^{2}_{y})^{3/2}}
\end{equation}
where $u_{x}$, $u_{xx}$, $u_{xy}$ and the rest are all approximately computed by introducing central finite differences into discrete form in practice.

However, calculating $\bar{K}$ in 3D by Eq.~\ref{h3D} is compute-intensive and time-consuming. Laplace operator or Laplacian is often used for edge detection by calculating the unmixed approximations of the second order derivatives with discrete and integer-valued convolution kernels with respect to each direction. It is safe to assume that the shape will have small change, then the first order derivatives will be approximated as small as zero, then Laplacian can be introduced to simplify the numerator of Eq.~\ref{h3D} by directly computing $u_{xx}$, $u_{yy}$, $u_{zz}$ by kernels to boost DNNs training speed. Our Laplacian-based 3D $curvature$ approximation can be expressed as follows,
\begin{equation}\label{cve3D1}
\bar{K} \approx \nabla^{2}=u^2_{xx}+u^2_{yy}+u^2_{zz}
\end{equation}
where $\nabla$ is the nabla operation, $\nabla^{2}$ is the Laplace operator. In practice, Laplacian can be estimated thus can be estimated by convolution. Our Laplacian-based 3D ACE loss function, denoted as Fast-ACE, has following advantages, 1) it is relatively inexpensive in terms of computations because the huge amount of work in 3D convolving operations could be computed by GPUs during the training steps. 2) the 3D \textit{mean curvature} of the shape can be driven efficiently at a minimum of computations loss from the operators. 
\section{Experiments}
\subsection{Experimental setting}
\textbf{Datasets: }In order to evaluate the performance of our ACE loss function, we have applied it to one natural and three biomedical image datasets: (1) 2D pavement crack images known as \textit{CrackTree200} \cite{zou2012cracktree}: CrackTree200 includes 206 pavement images of size 800 $\times$ 600 pixels with various types of cracks. 166 images were randomly selected for training, 20 images for validation and 20 images for testing. All the cases were down-sampled to the resolution of 448 $\times$ 448 pixels. (2) 2D Digital Retinal Images for Vessel Extraction (DRIVE) retina vessel \cite{staal2004ridge}: DRIVE contains 40 colour fundus images. Manual annotations by two experts are provided in DRIVE whereas the first one is chosen as the ground truth for performance evaluation. The 40 resized images (448 $\times$ 448) were divided into 50\% of images for training, 25\% for validation and the rest of images for testing. (3) 3D CT image of Pancreas from the Medical Segmentation Decathlon (MSD) \cite{simpson2019large}: This dataset consists of 281 3D abdominal CT images. We randomly selected 200 cases for the training, 20 for validation and 61 for testing. Following \cite{zhou2019prior}, we used the soft tissue CT window range of [-125, 275]. Then, we re-sampled all the cases to the resolution of 1.0 $\times$ 1.0 $\times$ 3.0 $mm^3$. Finally, we cropped all images centering at the pancreas region and normalized them to zero mean and unit variance. (4) 3D MR image of Left Atrium: A total of 100 3D gadolinium-enhanced MR imaging scans (GE-MRIs) from 2018 Atrial Segmentation Challenge \cite{xiong2018fully}. Following the experimental setting of \cite{yu2019uncertainty}, we used 16 cases for training, 10 cases for validation and 20 cases for evaluation. All cases were cropped centering at the heart region and normalized to zero mean and unit variance.


\textbf{Network architectures: }To investigate the robustness and generalizability of the ACE loss function, U-Net \cite{ronneberger2015u} and Context Encoder Network (CE-Net) \cite{gu2019net} are used as the 2D segmentation networks whilst 3D U-Net \cite{cciccek20163d} and V-Net \cite{milletari2016v} as our 3D segmentation networks. All the above networks are not pre-trained on any image datasets.

\textbf{Training and inference:} All the models were implemented by using Python 3.7 and PyTorch 1.4.0. All training experiments were done via one node of a cluster with sixteen 8-core Intel CPUs, 8 TESLA V100 GPUs and 1TB memory. The batch size and total training epochs were set as 8 and 600 respectively. All the models were trained by using the Adam optimizer. For a fair comparison of different loss functions, we searched the optimal learning rate in [$e^{-1}$, $e^{-2}$, $e^{-3}$, $e^{-4}$ ] for each loss function respectively based on the validation set \cite{karimi2019reducing}. During the training stage, we used the standard on-the-fly data augmentation methods to enlarge dataset and avoid over-fitting \cite{cciccek20163d}. In the inference phase, for a fair comparison we did not use any post-processing method to boost the performance.

\textbf{Evaluation metrics:} Two widely used metrics, Dice coefficient score (DICE) and the 95$^{th}$ percentile of Hausdorff Distance (HD$_{95}$) are used to quantitatively evaluate the segmentation results.
\subsection{Ablation study}
\begin{figure}[hbt]
	\centering
	\includegraphics[width=0.95\textwidth]{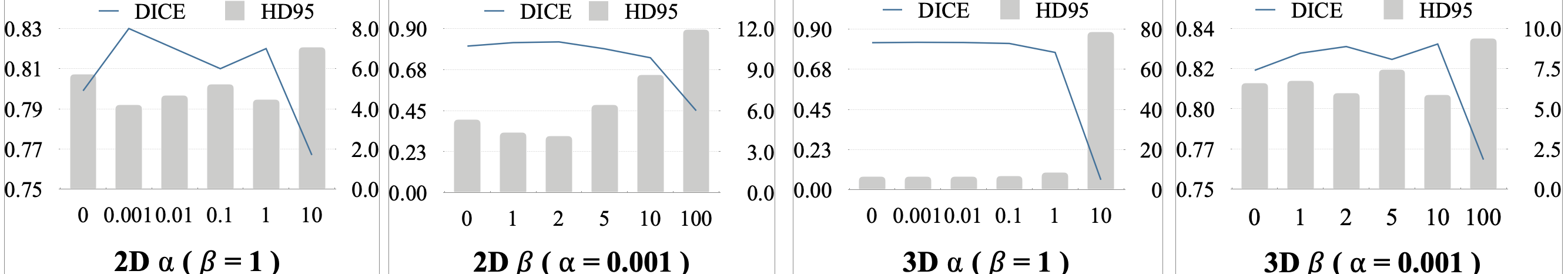}
	\caption{The trends of the segmentation performance when using U-Net with the ACE loss on the 2D DRIVE and 3D Pancreas validation set when $\alpha$ and $\beta$ are set to different values respectively.}
	\label{fig:onecol}
\end{figure}

We first investigated the optimal value of regularization weights $\alpha$ and $\beta$ in our proposed ACE loss in Eq.~\ref{cveloss} based on all validation datasets with 2D- and 3D- U-Net respectively. To investigate the individual impact of different $\alpha$ and $\beta$ values, we introduced variable-controlling method to perform this ablation study. Firstly, we fixed the $\beta$ to 1 to investigate the impact of $\alpha$ for ACE loss function performance. Then, we set the $\alpha$ to 0.001 in all experiments to investigate the impact of $\beta$ on model performance. Fig.~\ref{fig:onecol} shows the trends of the segmentation performance in terms of DICE and HD$_{95}$ on 2D DRIVE validation set and 3D Pancreas CT validation set when $\alpha$ and $\beta$ are with different values, respectively. It can be observed that increasing $\alpha$ from $0$ to 0.001 leads to an improved performance in the 2D $\alpha$ ablation experiment. When $\alpha$ is greater than $0.001$, the segmentation performance in 2D decreases gradually. In the 2D $\beta$ ablation experiment, the performance of $\beta$=1 is better than the performance of $\beta$=0 (the \textit{curvature} constant is not involved) in terms of HD$_{95}$. When $\beta $ is larger than $2$, the segmentation performance decreases significantly. The $\alpha$ and $\beta$ of ACE loss affect the 3D Pancreas segmentation performance in a similar way. Increasing the $\alpha$ from $0$ to 0.1 leads to an improved performance, when $\alpha$ is larger than 0.1, the segmentation performance worse rapidly. For the value of  $\beta$, it can be observed that the performance of $\beta$=1 is better than the performance of $\beta$=0 (the \textit{curvature} constant is not involved) in terms of DICE. Based on our observation from the above ablation study, our proposed ACE loss function has potential to be deployed in different DNNs-based image segmentation tasks by using well-chosen $\alpha$ in the range of [0.0001, 0.1] and the $\beta$ values in the structure-wised range of (0, 10]. It has the best performance on 2D DRIVE when $\alpha$ and $\beta$ are set to 0.001 and 2, respectively. For 2D \textit{CrackTree200}, $\alpha$ = 0.001 and $\beta$ = 0.7 have the best performance among other values. For the 3D Prances data, it has the best performance when $\alpha$ =  0.001 and $\beta$ = 10. For the 3D Left Atrial data, we also searched the optimal value of $\alpha$ and $\beta$ based on validation set ($\alpha$ =  0.001 and $\beta$=2). All the $\alpha$ and $\beta$ were set as optimal values in the following experiments.
\begin{table}[hb]
\centering
\caption{Quantitative 2D segmentation results (mean $\pm$ standard deviation) of our proposed ACE loss function compared to other loss functions. CE \cite{ronneberger2015u}, DC \cite{milletari2016v}, ClDC \cite{shit2020cldice}, AC \cite{chen2019learning} and our ACE loss function are evaluated on the CRACK200 and DRIVE datasets, respectively.}
\resizebox{\textwidth}{!}{%
\begin{tabular}{cccccccc}
\textbf{Dataset} & \textbf{Loss} & \textbf{Networks} & \textbf{DICE $\uparrow$} & \textbf{HD$_{95}$ $\downarrow$} & \textbf{Networks} & \textbf{DICE $\uparrow$} & \textbf{HD$_{95}$ $\downarrow$} \\ \hline
\multirow{5}{*}{CrackTree200} & CE & \multirow{5}{*}{U-Net} & 0.413$\pm$0.055 & 10.350$\pm$22.655 & \multirow{5}{*}{CE-Net} & 0.152$\pm$0.055 & 71.875$\pm$47.765 \\
 & DC &  & 0.571$\pm$0.046 & 7.107$\pm$26.964 &  & 0.594$\pm$0.042 & 7.624$\pm$26.874 \\
 & ClDC &  & 0.587$\pm$0.039 & 6.672$\pm$27.046 &  & 0.675$\pm$0.043 & 6.801$\pm$28.298 \\
 & AC &  & 0.640$\pm$0.039 & 6.362$\pm$27.159 &  & 0.664$\pm$0.040 & 6.149$\pm$27.947 \\
 & ACE &  & \textbf{0.675$\pm$0.034} & \textbf{5.742$\pm$28.425} &  & \textbf{0.683$\pm$0.041} & \textbf{5.902$\pm$28.281} \\ \hline
\multirow{5}{*}{DRIVE} & CE & \multirow{5}{*}{U-Net} & 0.734$\pm$0.046 & 9.350$\pm$3.837 & \multirow{5}{*}{CE-Net} & 0.723$\pm$0.038 & 12.033$\pm$3.332 \\
 & DC &  & 0.788$\pm$0.033 & 9.112$\pm$3.062 &  & 0.748$\pm$0.034 & 12.114$\pm$3.450 \\
 & ClDC &  & 0.796$\pm$0.032 & 8.241$\pm$3.394 &  & 0.782$\pm$0.027 & 11.568$\pm$3.643 \\
 & AC &  & 0.811$\pm$0.015 & 5.136$\pm$1.250 &  & 0.778$\pm$0.028 & 8.329$\pm$2.790 \\
 & ACE &  & \textbf{0.833$\pm$0.019} & \textbf{4.068$\pm$1.719} &  & \textbf{0.806$\pm$0.020} & \textbf{5.802$\pm$2.232} \\ \hline
\end{tabular}%
}
\label{comparisons2D}
\end{table}
\begin{figure}[!ht]
	\centering
	\includegraphics[width=1.0\linewidth, height=9cm]{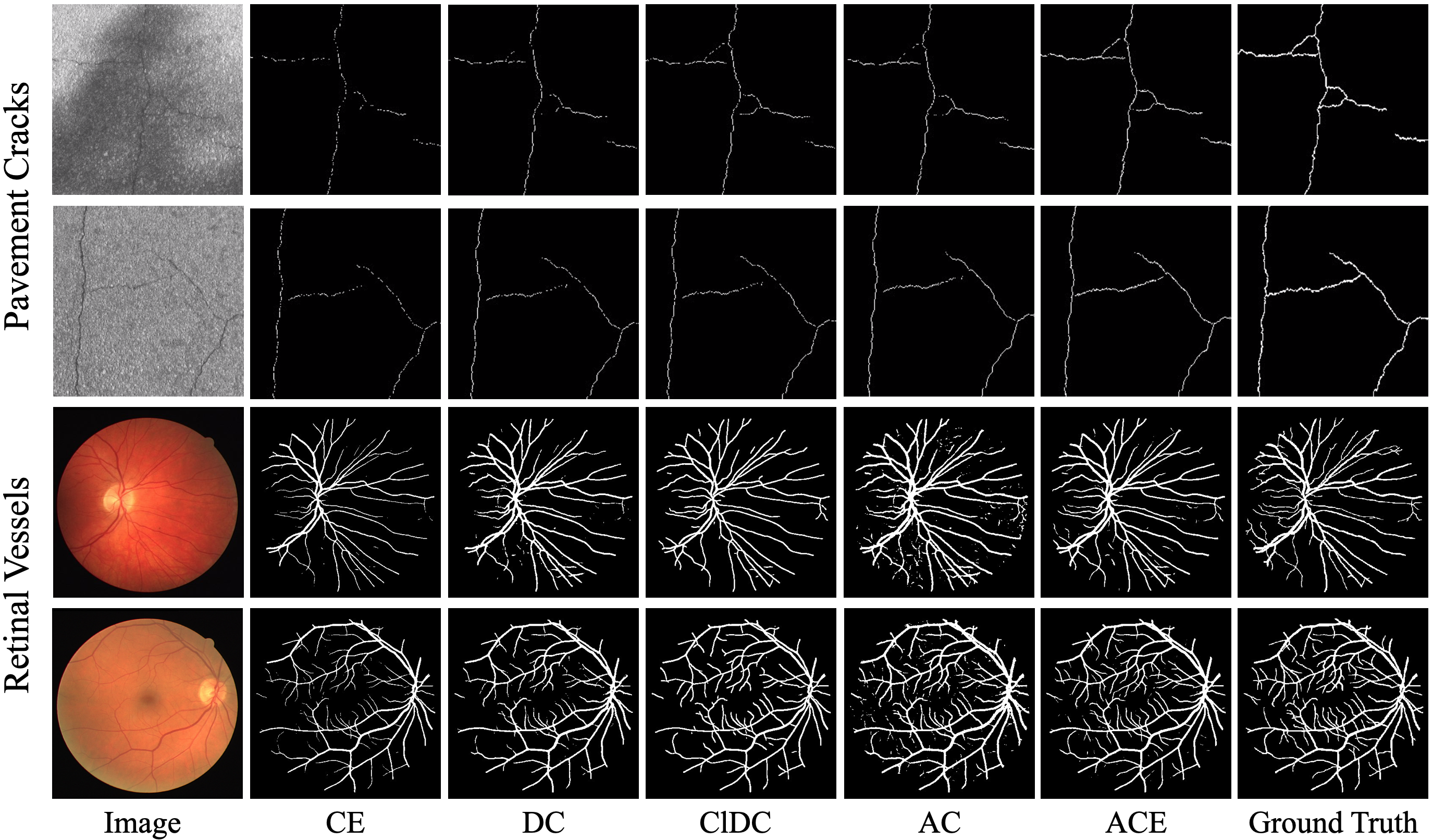}
	\caption{Example 2D segmentation results of our proposed ACE loss function compared to the other loss functions. The first and second row are pavement cracks segmentation results and the third and fourth row are retinal vessel segmentation results. From left to right are the original image, CE, DC, ClDC, AC, ACE loss functions and ground truth, respectively.}
	 \label{fig:sub1}
\end{figure}
\begin{table*}[t]
\centering
\caption{Quantitative comparison of CE \cite{ronneberger2015u}, DC \cite{milletari2016v}, HD \cite{karimi2019reducing}, AC \cite{chen2019learning}, our ACE and our Fast-ACE loss functions on 3D Pancreas CT and Left Atrial MRI, respectively. (mean $\pm$ standard deviation)}
\resizebox{\textwidth}{!}{%
\begin{tabular}{cccccccccc}

\textbf{Dataset} & \textbf{Loss} & \textbf{Networks} & \textbf{DICE $\uparrow$} & \textbf{HD$_{95}$ $\downarrow$} & \textbf{\begin{tabular}[c]{@{}c@{}}Time $\downarrow$\\ (s / epoch)\end{tabular}} & \textbf{Networks} & \textbf{DICE $\uparrow$} & \textbf{HD$_{95}$ $\downarrow$} & \textbf{\begin{tabular}[c]{@{}c@{}}Time $\downarrow$\\ (s / epoch)\end{tabular}} \\ \hline

 & { CE} &  & { 0.804$\pm$0.068} & 6.522$\pm$3.965 & 77.2 &  & { 0.796$\pm$ 0.081} & { 7.239$\pm$4.943} & 76.5 \\
 & { DC} &  & { 0.813$\pm$0.064} & 6.164$\pm$3.674 & 64.3 &  & { 0.791$\pm$0.085} & { 6.703$\pm$4.123} & 61.8 \\
 & { HD} &  & { 0.816$\pm$0.062} & 6.182$\pm$3.749 & 183.0 &  & { 0.816$\pm$0.078} & { 6.164$\pm$4.041} & 176.5 \\
 & { AC} &  & { 0.828$\pm$0.064} & 6.243$\pm$4.773 & \textbf{59.7} &  & { 0.819$\pm$0.096} & { 6.151$\pm$5.307} & \textbf{55.5} \\
 & { ACE} &  & { 0.835$\pm$0.059} & 5.521$\pm$3.298 & 78.1 &  & { 0.827$\pm$0.072} & { 6.124$\pm$4.494} & 75.2 \\
\multirow{-6}{*}{Pancreas} & Fast-ACE & \multirow{-6}{*}{{ 3D U-Net}} & \textbf{0.837$\pm$0.059} & \textbf{5.481$\pm$3.354} & 65.5 & \multirow{-6}{*}{{ V-Net}} & \textbf{0.832$\pm$0.070} & \textbf{6.013$\pm$5.314} & 61.5 \\ \hline
 
 & { CE} &  & { 0.861$\pm$0.043} & 14.659$\pm$6.243 & 3.2 &  & { 0.857$\pm$0.086} & { 14.846$\pm$10.835} & 3.4 \\
 
 & DC &  & 0.860$\pm$0.103 & 12.363$\pm$8.099 & 3.2 &  & 0.856$\pm$0.088 & 12.616$\pm$7.380 & \textbf{3.3} \\
 
 & HD &  & 0.879$\pm$0.084 & 11.609$\pm$8.163 & 8.2 &  & 0.865$\pm$0.056 & 17.358$\pm$17.138 & 8.4 \\
 
 & AC &  & 0.882$\pm$0.031 & 10.814$\pm$5.222 & \textbf{3.2} &  & 0.862$\pm$0.059 & 15.904$\pm$13.209 & 3.4 \\

 & ACE &  & 0.888$\pm$0.032 & 13.102$\pm$12.459 & 3.4 &  & 0.871$\pm$0.072 & 11.368$\pm$8.457 & 3.5 \\
\multirow{-6}{*}{Left Atrial} & Fast-ACE & \multirow{-6}{*}{{ 3D U-Net}} & \textbf{0.891$\pm$0.029} & \textbf{9.723$\pm$6.189} & 3.3 & \multirow{-6}{*}{{ V-Net}} & \textbf{0.882$\pm$0.049} & \textbf{9.057$\pm$3.474} & 3.4 \\ \hline
\end{tabular}
}
\label{comparison3D}
\end{table*}
\begin{figure}[t]
	\centering
	\includegraphics[width=1.0\linewidth]{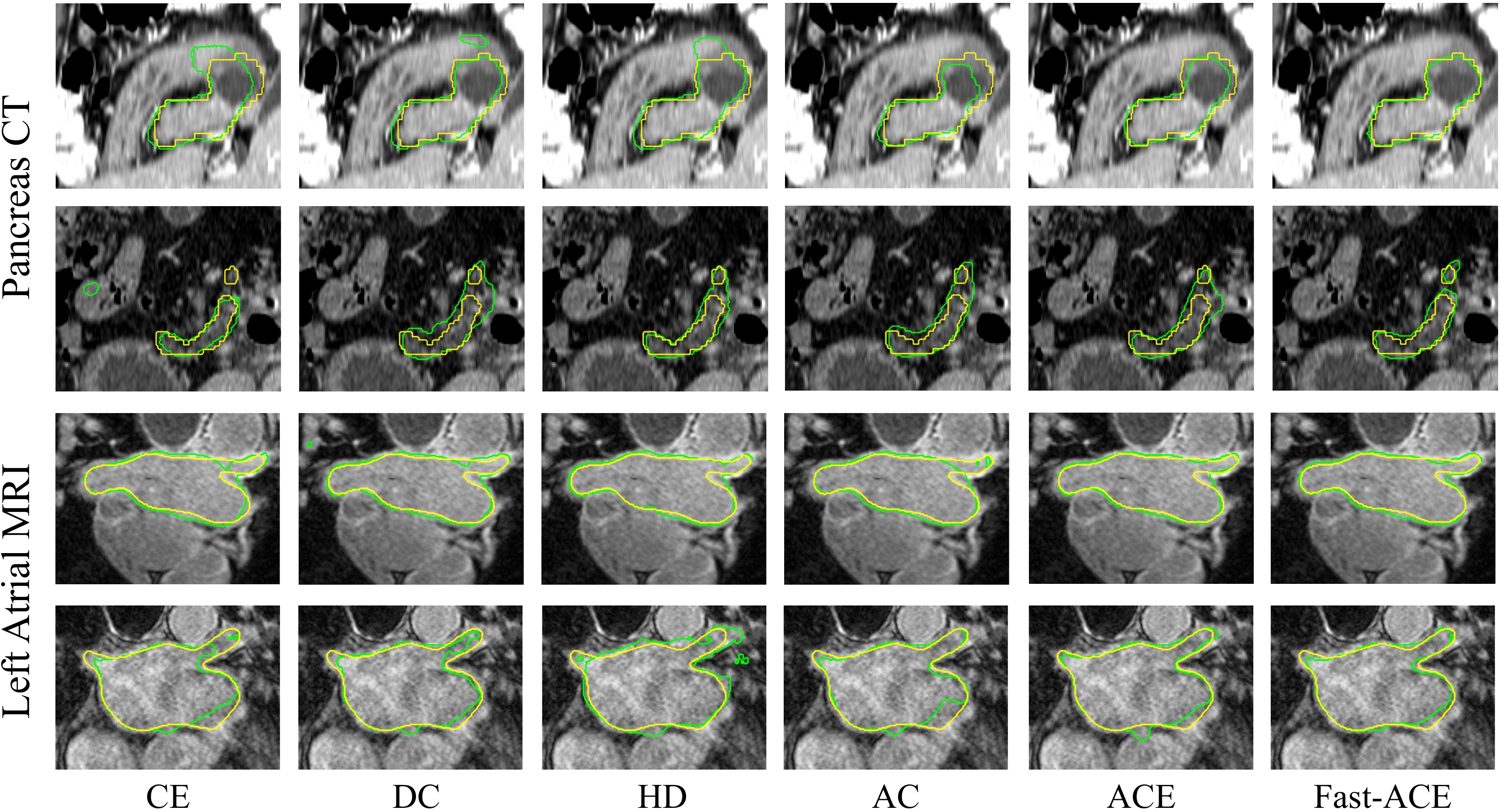}
	\caption{Example 3D segmentation results of our proposed ACE and Fast-ACE loss function compared to the other loss functions. The first and second row are pancreas segmentation results and the third and fourth row are left atrial segmentation results. The green and yellow contours denote the segmentation and the ground truth, respectively.}
	\label{fig:sub2}
\end{figure}
\subsection{Comparison to other loss functions}
We further compared the performance of the ACE loss with five widely used loss functions when appropriate: CE \cite{ronneberger2015u}, DC \cite{milletari2016v}, ClDC \cite{shit2020cldice}, AC \cite{chen2019learning}. We evaluated the extendibility and effectiveness of our ACE loss by training four different DNNs (U-Net and CE-Net for 2D curvilinear segmentation; 3D U-Net and V-Net for 3D organ segmentation).

\textbf{Curvilinear structure segmentation from 2D images: } We compared ACE loss with CE, DC, ClDC and AC functions. CE loss and DC loss are the most widely-used loss functions for image segmentation. ClDC loss and AC loss are proposed recently incorporating morphological skeletons and active contours respectively to enable DNNs focus on the objects geometric information. Table ~\ref{comparisons2D} presents quantitative comparisons of different loss functions on the two 2D datasets. For the DRIVE data, the ACE loss achieved better performance than all the other loss functions in terms of DICE of 0.833$\pm$0.019 (U-Net) and 0.806$\pm$0.020 (CE-Net), HD$_{95}$ of 4.068$\pm$1.719 pixels (U-Net) and 5.802$\pm$2.232 pixels (CE-Net), respectively. For the \textit{CRACK200} data, the ACE loss also performed better: DICE of 0.675$\pm$0.034 (U-Net) and 0.683$\pm$0.04 (CE-Net), HD$_{95}$ of 5.742$\pm$28.425 pixels (U-Net) and 5.902$\pm$28.281 pixels (CE-Net), respectively. Figure \ref{fig:sub1} shows four segmentation results of using U-Net with five different loss functions, which highlights the ACE loss is able to preserve curvilinear structure connectedness. 

\textbf{Segmentation of 3D CT and MR images:} We compared our proposed ACE and Fast-ACE loss functions with CE, DC, HD and AC loss functions on pancreas CT and left atrial MRI datasets and results are presented in Table \ref{comparison3D}. For the pancreas segmentation, the Fast-ACE loss with 3D U-Net and V-Net outperform existing loss functions in terms of DICE of 0.832$\pm$0.070 (U-Net) and 0.837$\pm$0.059(V-Net), HD$_{95}$ of 6.013$\pm$5.314 voxels (U-Net) and 5.481$\pm$3.354 voxels (V-Net) respectively. For the  left atrial segmentation, the Fast-ACE loss with 3D U-Net and V-Net achieved the best performance on left atrial segmentation in term of DICE of 0.891$\pm$0.029, 0.882$\pm$0.049 and HD$_{95}$ of 9.723$\pm$6.189 voxels and 9.057$\pm$3.474 voxels respectively. It can be observed that both the ACE and Fast-ACE loss do not rely on specific-designed network and thus has very good generalizability. For the computing time during training stage, the Fast-ACE loss with V-Net spend 61.5s and 3.4 s per epoch on the pancreas and left atrial dataset respectively, shorter than ACE loss with V-Net in 75.2s and 3.5s per epoch. In Table \ref{comparison3D}, it can be observed that both the ACE loss and Fast-ACE loss used less than half of the time of the HD loss and are comparable with mainstream loss functions. There is almost no different in the prediction for the same network as the loss function will not be used. Figure \ref{fig:sub2} presents some segmentation results of different methods on the two datasets (all results obtained by V-Net). It can be observed that the segmentation results of the proposed ACE loss and Fast-ACE loss are more accurate compared with the other existing methods. 
\begin{table}[!ht]
\centering
\caption{The comparison of U-Net with our proposed ACE loss function and other SOTA approaches on DRIVE and Left Atrial MRI, respectively.}
\label{vsothers}
\resizebox{\textwidth}{!}{%
\begin{tabular}{ccccccc}
\textbf{Dataset} & \textbf{Method} & \textbf{Sensitivity $\uparrow$} & \textbf{AUC $\uparrow$} & \textbf{DICE $\uparrow$} & \textbf{HD$_{95}$ $\downarrow$} & \textbf{\begin{tabular}[c]{@{}c@{}}Time $\downarrow$\\ (s / epoch)\end{tabular}} \\ \hline
\multirow{5}{*}{DRIVE} & Vega et al. \cite{vega2015retinal} & 0.740 & - & 0.690 & - & - \\
 & Maninis et al.  \cite{maninis2016deep} & - & - & 0.820 & - & - \\
 & Zhao et al. \cite{zhao2017automatic} & 0.774 & 0.975 & 0.793 & - & - \\
 & Gu et al. \cite{gu2019net} & 0.831 & 0.978 & - & - & - \\
 & 2D U-Net + ACE (ours) & 0.805 & 0.979 & 0.831 & 4.07 & 1.8 \\ \hline
\multirow{4}{*}{Left Atrial} & Vanilla V-Net \cite{yu2019uncertainty} & - & - & 0.841 & 17.93 & - \\
 & Bayesian V-Net \cite{yu2019uncertainty} & - & - & 0.860 & 14.26 & - \\
 & Yu et al. \cite{yu2019uncertainty} & - & - & 0.882 & 11.40 & 14.2 \\
 & 3D U-Net + Fast-ACE (ours) & - & - & 0.891 & 9.72 & 3.3 \\ \hline
\end{tabular}%
}
\end{table}

We also compared our best segmentation performance with state-of-the-art approaches on DRIVE and left atrial datasets as shown in Table \ref{vsothers}. For DRIVE data, our best result in terms of DICE of 0.833 and area under the receiver operating characteristics curve (AUC) of 0.979 by a 2D U-Net with our ACE loss function is higher than the other state-of-the-art results of retinal vessels segmentation. For left atrial segmentation, our best segmentation result (DICE = 0.891, HD = 9.72) by a 3D U-net with Fast-ACE loss function is better than to the results by other state-of-the-art approaches.

\section{Discussion \& Conclusion}
In this work, we proposed and implemented a new ACE loss for DNNs-based end-to-end image segmentation. Being a differentiable loss function, the ACE loss comprises elastica (curvature and length) and region constraints to achieve more robust and accurate segmentation. Compare to standard ACMs which require iterations to in solving PDEs for each single image, the use of supervised DNNs will hugely reduce the computational time on segmenting new images after the training. We found that the regularization weight $\alpha $ in the elastica constraints can be fixed to 0.001 for different image segmentation tasks. The regularization weight $\beta $ is more sensitive to different images and objects structures. For instance, for curvilinear or tubular structures image segmentation tasks, a $\beta $ (0 < $\beta < 2 $) has a better segmentation results whilst a $\beta $ ($2  < \beta < 10 $) for non-tubular structures. In the future, developing a novel model to learn different regularization weights in elastica will be desirable~\cite{schoenemann2011elastic}. Notably, from the quantitative comparison of segmentation performance between AC and our ACE loss, we observe that our elastica constraint is more effective to use the geometrical information to constrain segmentation process than the length constraint only of the AC loss, which leads to improved segmentation results in terms of DICE and HD$_{95}$ with four different networks on four different image datasets. With respects to the learning efficiency, when the number of the training image set is relatively large (e.g. the Pancreas dataset), the proposed Fast-ACE loss can maintain similar results but improve the training efficiency by about 20\% (this improvement may not be significant for small dataset). 

In summary, we introduced a novel ACE loss function for effective and accurate segmentation tasks by supervised DNN approaches. The advantage of this new loss function is that it can seamlessly integrate the geometrical information (e.g. curvature and length of the target shape) with region similarity thus leading to more accurate and reliable segmentation. We introduced mean curvature as a more precise image prior to represent curvature in our ACE loss. Based on the definition of mean curvature, we propose a fast 3D solution Fast-ACE to speed up training process for 3D image segmentation. We applied both the ACE and Fast-ACE to four datasets (two 2D and 3D each) and the results showed that they outperforms state-of-the-art loss functions. It is believed that this new loss function will be readily applied to other computer vision tasks.

\bibliographystyle{unsrt}  
\bibliography{arxiv.bbl}
\end{document}